\documentclass[nofootinbib,aps,pra,reprint,superscriptaddress,amsmath,amsfonts,amssymb]{revtex4-1}

\usepackage{todonotes}
\usepackage{graphicx}
\usepackage{bm}
\usepackage{siunitx}
\usepackage[hidelinks]{hyperref}
\usepackage[bottom]{footmisc}
\usepackage[capitalise]{cleveref}
\usepackage{lipsum}
\usepackage[version=4]{mhchem}

\renewcommand{\vec}[1]{\mathbf{#1}}
\newcommand{\bk}[1]{|#1\rangle}
\newcommand{\ket}[1]{|#1\rangle}

\newcommand{\G}{\Gamma}
\newcommand{\ps}{\psi}
\newcommand{\intra}[1]{t_{#1}}
\newcommand{\inter}[1]{s_{#1}}
\newcommand{\av}[1]{\vec{a}_{#1}}
\newcommand{\expt}[2][]{e^{#1 i\vec{k}\cdot #2}}

\newcommand{\ta}{\theta}
\newcommand{\oo}{\omega}
\newcommand{\D}{\Delta}

\newcommand{\ii}{\mathrm{i}}
\newcommand{\ee}{\mathrm{e}}
\newcommand{\degr}{^{\mathrm{o}}}

\newcommand{\etal}{\textit{et al}.~}

\begin{document}
	
	\title{Direct Observation of Topological Edge States in Silicon Photonic Crystals: \\Spin, Dispersion, and Chiral Routing}
	
	\author{Nikhil Parappurath}
	\affiliation{Center for Nanophotonics, AMOLF, Science Park 104, 1098 XG Amsterdam, The Netherlands}
	\author{Filippo Alpeggiani}
	\affiliation{Department of Quantum Nanoscience, Kavli Institute of Nanoscience, Delft University of Technology, 2600 GA Delft, The Netherlands}
	\author{L. Kuipers}
	\affiliation{Department of Quantum Nanoscience, Kavli Institute of Nanoscience, Delft University of Technology, 2600 GA Delft, The Netherlands}
	\author{Ewold Verhagen}
    \email[Corresponding author: ]{verhagen@amolf.nl}
	\affiliation{Center for Nanophotonics, AMOLF, Science Park 104, 1098 XG Amsterdam, The Netherlands}
	
	\date{\today}

	\begin{abstract}
	
		Topological protection in photonics offers new prospects for guiding and manipulating classical and quantum information. The mechanism of spin-orbit coupling promises the emergence of edge states that are helical; exhibiting unidirectional propagation that is topologically protected against back-scattering. We directly observe the topological states of a photonic analogue of electronic materials exhibiting the quantum Spin Hall effect, living at the interface between two silicon photonic crystals with different topological order. Through the far-field radiation that is inherent to the states' existence we characterize their properties, including linear dispersion and low loss. Importantly, we find that the edge state pseudospin is encoded in unique circular far-field polarization and linked to unidirectional propagation, thus revealing a signature of the underlying photonic spin-orbit coupling. We use this connection to selectively excite different edge states with polarized light, and directly visualize their routing along sharp chiral waveguide junctions.
	\end{abstract}
	\maketitle

	The concept of topology has proven immensely powerful in physics, describing new phases of matter with unique properties. The connection of the quantum Hall effect \cite{von_klitzing_quantized_1986} to band structure topology explained the emergence of topologically protected unidirectional transport of electrons. These states exist at the edge of two-dimensional systems subject to an external magnetic field that breaks time-reversal (TR) symmetry. The prediction \cite{kane_quantum_2005} and observation \cite{bernevig_quantum_2006,konig_quantum_2007} of the quantum spin-Hall effect (QSHE) unlocked a new door in the field of topological physics. This effect  relies on spin-orbit coupling instead of an external magnetic field, causing spin-up and -down electrons to propagate in opposite directions in states that are protected by TR symmetry. There has been a recent surge in attempts to implement topological protection in the acoustic, mechanical, microwave and optical domains \cite{wang_observation_2009,khanikaev_photonic_2013,rechtsman_photonic_2013,chen_experimental_2014,bellec_manipulation_2014,slobozhanyuk_subwavelength_2015,fleury_floquet_2016,he_acoustic_2016,cheng_robust_2016,milicevic_orbital_2017,noh_observation_2018,lumer_observation_2018,stutzer_photonic_2018,brendel_snowflake_2018,ozawa_topological_2018}, owing to the application potential of robust transport. A particular opportunity is provided by photonic spin-orbit coupling \cite{bliokh_quantum_2015,sala_spin-orbit_2015}.
	Photonic analogues of QSHE were realized using arrayed ring resonators, where the helicity of propagation in the ring takes the role of (pseudo)spin \cite{hafezi_imaging_2013,mittal_topologically_2014,mittal_measurement_2016,mittal_topological_2018-1,bandres_topological_2018}. Recently, QSHE was also predicted to occur in photonic crystals with special symmetries, with edge state propagation direction linked to local circular Poynting vector flows \cite{wu_scheme_2015,barik_two-dimensionally_2016,anderson_unidirectional_2017}. Such states were observed in the microwave domain \cite{yves_crystalline_2017,yang_visualization_2018}, and coupled to spin-polarized quantum dots \cite{barik_topological_2018}. Realizing topological photonic states in the telecom domain is especially promising in view of exciting possibilities like nanoscale routing, resilience to disorder, one-way transport and robust light emission in integrated photonic chips. In contrast to valley-Hall topological photonic crystals \cite{ma_all-si_2016, shalaev_robust_2018, he_silicon--insulator_2018} that operate below the light line, topological photonic crystals employing QSHE offer the possibility to access their properties via far-field radiation \cite{gorlach_far-field_2018}.

	\begin{figure*}
		\centering
		\includegraphics{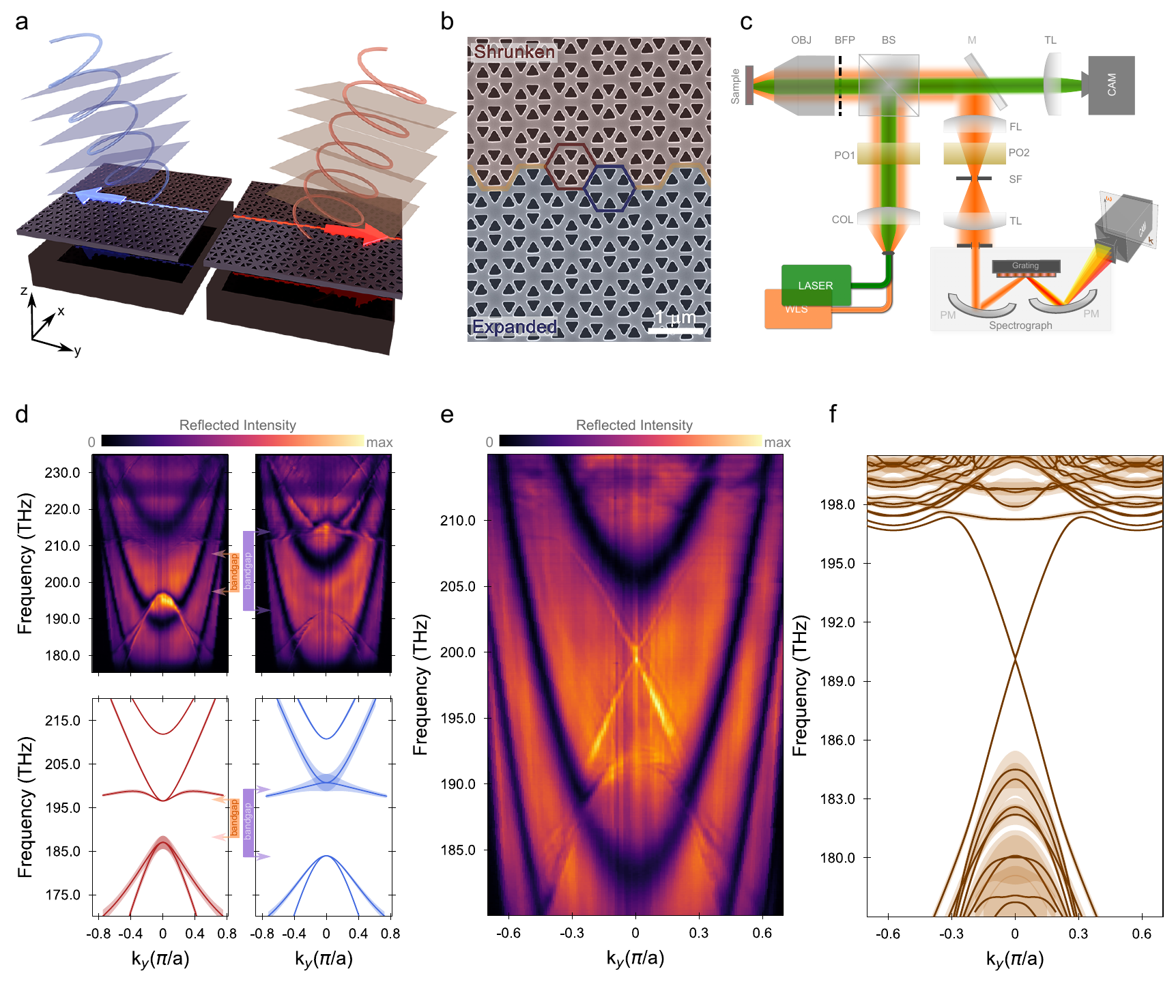}
		\caption{a. Schematic representation of spin-orbit coupling in topological photonic crystals: Chirality of the radiation field is connected to propagation direction of states at edges between photonic crystals with different topological bandstructure. b. A scanning electron micrograph of the fabricated sample. The yellow line represents the arm-chair like edge between two pseudo-colored regions of differing topology. Unit cells on either side are highlighted by red and blue hexagons. c. Schematic diagram of the experimental setup. The setup can operate in two modes: In the Fourier-space spectroscopy mode (orange path), a Fourier Lens (FL) images the back-focal plane (BFP) onto the spectrometer slit using a tube lens (TL). A spatial filter (SF) chooses the area from which light is collected. Quarter wave plates together with linear polarizers are used to define polarization of input light (PO1) and for polarimetry measurements in the reflected path (PO2). In the real-space imaging mode (green path), the sample plane is imaged onto the camera using another TL without FL and SF. See Methods for more details and abbreviations. d. Measured (top panel) and calculated (bottom panel) dispersion of shrunken (left) and expanded (right) lattices for diagonally polarized incidence. The measured reflection intensity is normalized with respect to the maximum pixel value among the images. Linewidths of the calculated modes, scaled by a factor of 5, are shown as shaded regions in the bottom panel. e, f. Measured and calculated dispersion of the edge states. Calculated linewidths are shown as shaded region in f.}
		\label{Fig-first}
	\end{figure*}
	
	Here we directly observe topological photonic states at telecom wavelengths in photonic crystals in silicon-on-insulator (SOI) technology. Through angular spectroscopy and polarimetry, we retrieve the intriguing properties of not only the bulk states of crystals with different topological order, but also of the edge states that appear at their interfaces. We determine their dispersion, and reveal that the radiation of the topological states carries their unique nature: The states' pseudospin is encoded in the circular polarization of their far field, and one-to-one related to the unidirectional propagation (Fig.~\ref{Fig-first}(a)). Through this connection, we selectively excite modes in opposite directions and map their propagation in real space. We show that the measured propagation quality factors of $\sim$450 are intrinsically limited by the very radiation we observe, and experimentally observe an anti-crossing between unidirectional modes due to symmetry breaking at the crystal edge. A tight-binding model fitted to the measured bulk state dispersion correctly predicts the magnitude of this spin-spin scattering. Finally, we show how the pseudospin in combination with a chiral waveguide crossing can be used to effectively route optical information.

	As photonic systems do not feature the Kramers doubling that guarantees the existence of electronic states of opposite spin in crystals, we follow a symmetry-based approach to topological photonics outlined by Wu and Hu \cite{wu_scheme_2015} and Barik \etal\cite{barik_two-dimensionally_2016}. The scheme starts from a photonic crystal with scattering sites arranged in a honeycomb lattice. Viewing such crystals as a triangular lattice of hexagonal unit cells with six sites each, they exhibit a doubly degenerate Dirac cone at the $\G$ point. We fabricate photonic crystals with hexagonal unit cells containing equilateral triangular holes in a silicon membrane through electron beam lithography and reactive ion and wet etching of a silicon-on-insulator substrate. Figure \ref{Fig-first}b shows a pseudo-colored scanning electron micrograph in top view. 
	
	Preserving the $C_6$ crystal symmetry, each hexagonal cluster is deformed either by concentrically shifting the holes inwards (`shrunken lattice', top of Fig.~\ref{Fig-first}b) or outwards (`expanded lattice', bottom of Fig.~\ref{Fig-first}b), without altering unit cell size. Such deformations open a gap at the Dirac point, but qualitatively in different ways. A continuous deformation of the lattice from shrunken to expanded is necessarily accompanied by a closing of the bandgap, akin to the behavior in a one-dimensional Su-Schrieffer-Haeger model \cite{su_solitons_1979,slobozhanyuk_subwavelength_2015,asboth_short_2016,st-jean_lasing_2017}. The shrunken and expanded lattices are then associated with two different band structure topologies analogous to a $\mathbb{Z}_2$ topological insulator. A band inversion takes place across the closing of the gap, meaning that the nature of the states at the top and bottom band edges is opposite for shrunken and expanded crystals. The states of the bottom (top) bands in the shrunken (expanded) lattice, characterized by the out-of-plane magnetic field $H_z$, resemble $p$-orbitals, whereas the top (bottom) bands are $d$-like \cite{wu_scheme_2015,barik_two-dimensionally_2016}. Bulk-edge correspondence guarantees the existence of topologically protected states at the edge of two domains of different topology, such as the armchair interface in Fig.~\ref{Fig-first}b.
	
	The unit cell deformation that opens the band gap also couples the TE-like states in the silicon slab to far-field radiation. As both the band edges and edge states appear close to the $\G$ point, they are naturally phase-matched to free-space radiation \cite{gorlach_far-field_2018}. In either expanded or shrunken lattices the fundamental Bloch harmonic of the photonic crystal eigenmodes can carry finite weight. We can thus use this radiation field to characterize the energy, momentum, localization, and polarization of the states of the crystals and their edges.
	
	\section*{Results and discussion}
	
	\subsection*{Bulk and edge state dispersion}
	We use the reflectometry setup outlined in Fig.~\ref{Fig-first}c to directly measure the photonic crystal band dispersion in reciprocal space. Focused, broadband incident light excites the crystal modes. Directly reflected and re-radiated light is collected from an area of $\sim 30 \times 102~\si{\micro\meter}$, restricted by a spatial filter (see Methods). By imaging the objective back focal plane onto the entrance slit of a spectrograph \cite{taminiau_quantifying_2012}, we record two-dimensional images of frequency versus $k_y$ wavevector of reflected light with $k_x=0$ on an IR camera at the spectrograph output.
	
	The top panels in Fig.~\ref{Fig-first}d show the measured reflection as a function of frequency and $k_y$ from shrunken and expanded lattice crystals for diagonally polarized input. The spectrally broad fringes are due to Fabry-P\'erot-like reflections between the suspended membrane and the substrate (see Supplementary Fig.~S1). We observe sharper resonant features interfering with this direct reflection background to produce bands with Fano lineshapes. The dispersion of these states corresponds well to the calculated photonic crystal band diagrams in the bottom of Fig.~\ref{Fig-first}d for both lattices, apart from an overall $\sim10~\si{\tera\hertz}$ frequency offset that is likely related to deviations of fabricated hole size and shape from designed geometry. Band gaps are visible around optical frequencies of $\sim200~\si{\tera\hertz}$, with widths of $9~\si{\tera\hertz}$ ($\D\oo/\oo_0\approx4.4\%$) for the shrunken lattice and $18~\si{\tera\hertz}$ ($\D\oo/\oo_0\approx8.9\%$) for the expanded lattice. These measured bandgaps are in good agreement with the calculated relative gaps of $4.7\%$ and $7.8\%$, respectively. 
	
	Figure~\ref{Fig-first}d shows the band inversion between shrunken and expanded lattices reflection, manifested in the leakiness of their upper and lower bands \cite{gorlach_far-field_2018}. The $p$-orbitals of $H_z$ associated with the bottom (top) band edge of the shrunken (expanded) lattice mean that their in-plane electric field distributions are even with respect to the center of the unit cell. As such, these states can couple to radiation normal to the slab, whereas the $d$-orbital states are quadrupolar and dark. We observe the bright spectral features of the $p$-orbital states at opposite sides of the gap for the two lattices, accompanied with linewidth reduction of the other edge. The latter is confirmed by the mode calculations in the bottom panels of Fig.~\ref{Fig-first}d, where the shading widths are proportional to the imaginary part of the eigenmode frequency. These radiation properties thus confirm that the two lattices feature band gaps of different topological character.
	
	Figure~\ref{Fig-first}e shows the band diagram measured in reflection when exciting the armchair-like edge between regions of expanded and shrunken lattices with a diagonally polarized focused beam. 
	We observe two states that close the bandgap with opposite velocities, crossing at the $\G$ point. They exhibit the characteristic linear dispersion of edge states in QSHE topological insulators, with a measured group velocity of $\sim c/8.3$ that is uniform over most of the effective band gap. The measured band diagrams correspond well to the calculated modes, shown in Fig.~\ref{Fig-first}f.
	\begin{figure}[!h]
		\centering
		\includegraphics[width=1\columnwidth]{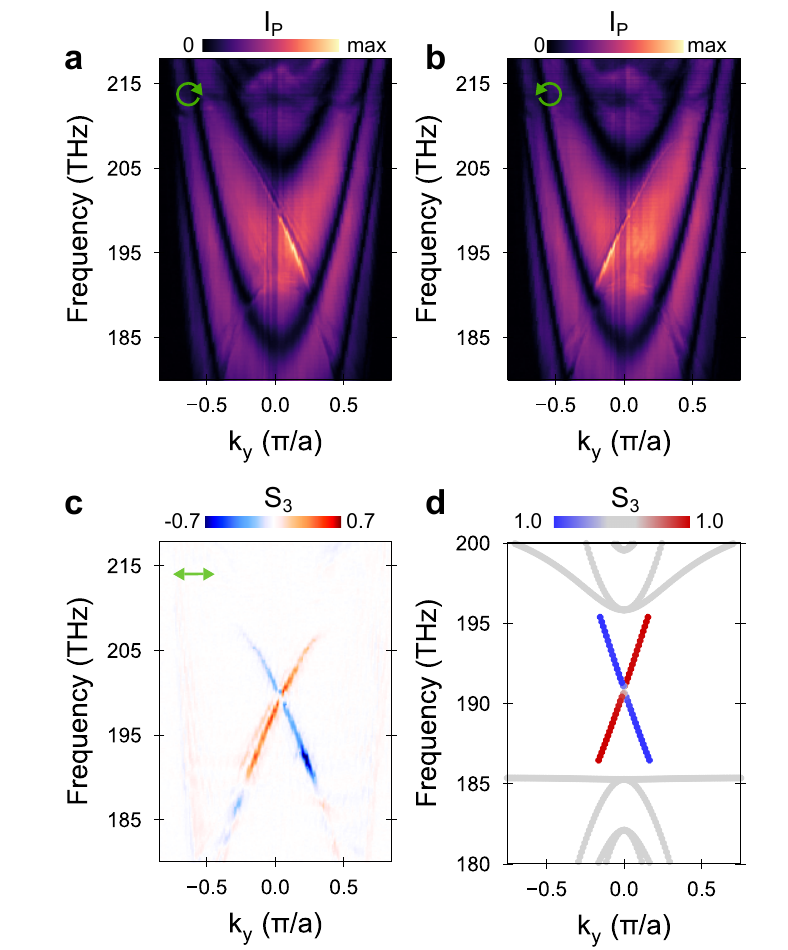}
		\caption{Polarimetry results. \textbf{a, b}. Total reflected intensity of the polarized field ($I_P$) as a function of frequency and wavevector $k_y$ when the edge is excited with right- (a) and left- (b) circularly polarized light. Images are normalized with respect to corresponding maximum pixel values. \textbf{c}. Measured circular polarization intensity (normalized Stokes parameter $S_3$) when the edge is excited with light that is linearly polarized along $y$. A center block is placed at the real space plane to spatially filter background reflections (see Methods). d. Normalized $S_3$ for the edge states as predicted by the tight-binding model.}
		\label{Fig-polarimetry}
	\end{figure}
	
	\subsection*{Detecting pseudospin}
	We then turn to study the polarization properties of these edge states, modifying the setup to add the functionality of polarimetry. This allows quantifying the polarization of the reflected field in terms of its Stokes parameters~\cite{born_principles_1999,schaefer_measuring_2007}. We first excite the edge with circularly polarized (CP) light of different handedness and plot the intensity of the polarized component of the reflected field, $I_P = \sqrt{S_1'^2+S_2'^2+S_3'^2}$, where $S_i$ are unnormalized Stokes parameters, in Figs.~\ref{Fig-polarimetry}a and b. One recognizes that the two edge states of positive or negative group velocity are excited selectively with near-perfect contrast using either left or right CP inputs, respectively.
	
	This observed helical nature of the edge states' far field can be understood by considering how these states originate from the $p$- and $d$-like orbitals in the infinite lattice unit cells. At the $\G$ point, the bands are degenerate. In the $H_z$ field basis of $\ket{p_\pm}=(\ket{p_x}\pm\ii \ket{p_y})/\sqrt{2}$ and $\ket{d_\pm}=(\ket{d_{(x^2-y^2)}}\pm\ii \ket{d_{(xy)}})/\sqrt{2}$, a tight-binding model that describes coupling between the sites results in a $4\times4$ Hamiltonian in the approximation of nearest-neighbor coupling and linear expansion in $k_x$ and $k_y$ that is block-diagonal, meaning that only states of the same pseudospin $\pm$ are coupled~\cite{wu_scheme_2015}. As a result, the edge states that connect the top and bottom bands must be associated with a single pseudospin, admixing either $\ket{p_+}$ and $\ket{d_+}$ or $\ket{p_-}$ and $\ket{d_-}$. Since the $p$-like component is predominantly responsible for the plane wave emitted to the far-field, one can see that the fundamental harmonic of the Bloch mode must be circularly polarized, with handedness equal to the pseudospin.  
	
	This radiation and its polarization can be quantified in the tight-binding model by calculating the total in-plane dipole moment inside the unit cell $\vec{p} \propto \sum_{i=1}^6 \hat{\vec{\theta}}_i \psi_i$, where $\hat{\theta}_i$ is the azimuthal unit vector of the $i^{th}$ site and $\ket{\psi}$ is a six-dimensional scalar ``wave function'' with components $\psi_{i}$ corresponding to $H_z$ at site $i$. 
	In order to also correctly predict the polarization of the bulk bands, we need to extend the tight-binding model beyond the nearest-neighbour approximation, which predicts degenerate bulk bands over the entire reciprocal space~\cite{barik_topological_2018}. In contrast, we observe splitting of both upper and lower bands away from the $\G$ point in experiment and finite-element simulation, with linear polarization (see Supplementary Fig.~S2). By introducing additional coupling terms up to third order in the tight binding model (see Supplementary Figure S3), we retrieve the observed lifting of degeneracy, allowing the prediction of the polarization of each observed state.
	
	To assess the polarization of the edge and bulk states experimentally, we excite the edge with light that is horizontally polarized, and perform full polarization tomography of the emitted radiation as a function of frequency and wavevector along the edge. To isolate the fields of the photonic crystal modes, the spatial filter is modified to block direct reflection from the slab and the substrate below (see Methods). Figure~\ref{Fig-polarimetry}c displays the measured Stokes parameter $S_3$ --- which quantifies the detected circularly polarized intensity --- normalized to the maximum value of total intensity of the polarized component. We see that only the edge states exhibit a significant degree of CP. Their opposing helicity is evident: The emission of the negative pseudospin state with negative group velocity is mostly left CP and that of the positive pseudospin state with positive group velocity is mostly right CP. In other words, we directly observe the pseudospin of the topological edge states, and the experiments reveal a clear signature of the photonic spin-orbit coupling that lies at the root of the QSHE. 
	
	\begin{figure}[!h]
		\centering
		\includegraphics[width=1\columnwidth]{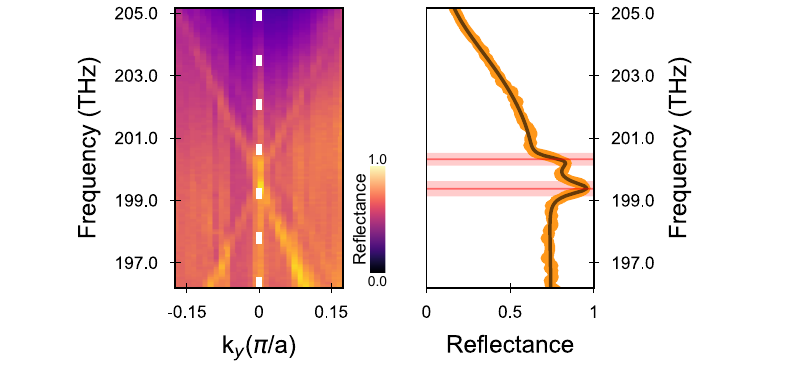}
		\caption{Quantifying spin-spin scattering in the system. a. Measured reflectance showing dispersion of edge states as a function of wavevector $k_y$. b. Measured intensity along the cross-cut indicated by the white line in a (orange data). The black line shows a fit with a set of complex Lorentzians (see Methods). Red lines and shaded areas correspond to the extracted resonance frequencies and linewidths.}
		\label{Fig-crosscut}
	\end{figure}
	
	\subsection*{Symmetry breaking and spin-spin scattering}
	The measured far-field polarization is in good agreement with the results of the tight-binding model, whose parameters are fitted to the observed frequencies. We find that, also with the inclusion of the higher-order coupling terms, the average dipole moment remains CP. This behavior is illustrated in Fig.~\ref{Fig-polarimetry}d which shows the Stokes parameter $S_3$ calculated by our theoretical model for edge as well as bulk modes. 
	
	Interestingly, the inclusion of higher-order coupling terms leads to the prediction of a small anticrossing between the two edge states, as can be seen in Fig.~\ref{Fig-polarimetry}d. Topological protection in these systems is linked to a `pseudo' time-reversal symmetry, which needs the preservation of $C_6$ symmetry while deforming the honeycomb lattice unit cells~\cite{lu_topological_2014}. However, at the interface between two deformed lattices, $C_6$ symmetry breaking is unavoidable. This results in photonic spin-spin scattering --- coupling of the counter-propagating edge states that opens a small gap around the $\G$ point~\cite{xu_accidental_2016}. We see this also in experiment as a discontinuity around 200 THz in Fig.~\ref{Fig-polarimetry}a and b. The amount of spin-spin scattering and thus the intrinsic limit to topological protection in photonic crystals can be recognized from this gap. 
	
	\begin{figure*}
		\centering
		\includegraphics[width=1\textwidth]{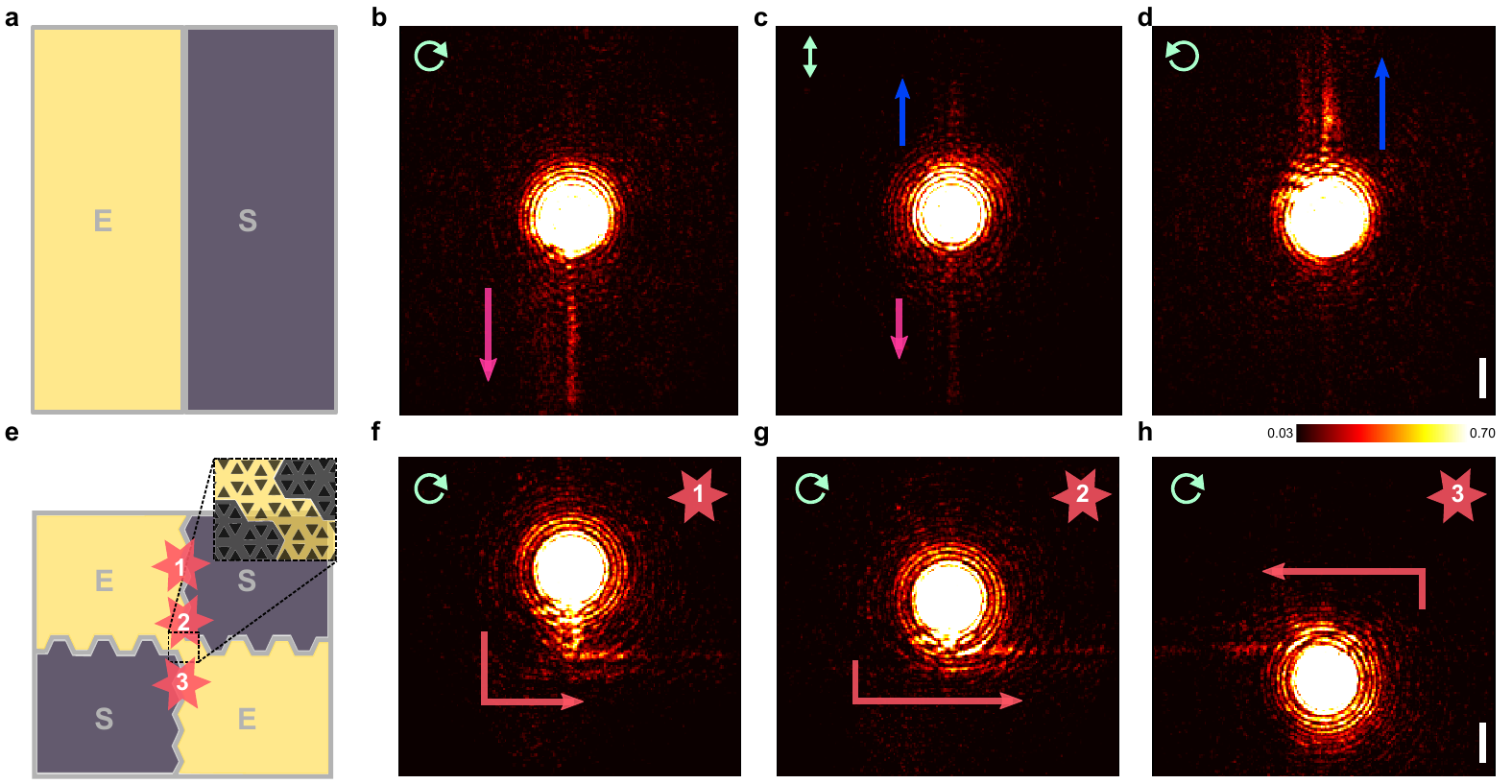}
		\caption{Real-space images of edge state propagation. a. A schematic diagram showing the edge between expanded (E) and shrunken (S) lattices where a focused single wavelength laser excites the sample. b-d. Real space camera images when the incident light is right circularly, linearly and left circularly polarized, respectively. Images are normalized to the maximum pixel value among the three. e-h. Routing of light around a topologically protected chiral junction. e. A schematic diagram showing a junction formed by four topological edges. Inset shows a zoomed-in image of the chiral junction. Real-space images f, g and h show light propagation when the sample is excited by a right circularly polarized single wavelength laser at positions 1, 2 and 3 marked in e. All three images f-h are normalized to the maximum pixel value among the three. Scalebars correspond to $10~\si{\micro\meter}.$}
		\label{Fig-realspace}
	\end{figure*}
	
	Figure~\ref{Fig-crosscut}a shows reflected intensity normalized to that from a silver mirror around the $\G$ point for horizontally polarized input, showing the anticrossing. The reflectance at $k_y=0$ (Fig.~\ref{Fig-crosscut}b) is fitted with two complex Lorentzians together with a slowly varying background (see Methods). The edge states are split by a $1.0~\si{\tera\hertz}$ gap. This compares well to the $0.6~\si{\tera\hertz}$ gap predicted by the tight-binding model, whose parameters were obtained from fitting to bulk simulation data, independently of the gap. We do note that the bandgap predicted by finite-element calculations was somewhat smaller ($\sim 0.1~\si{\tera\hertz}$) for the armchair edge. Further studies should determine whether spin-spin scattering is practically limited by fundamental symmetry breaking at the interface, or by other factors including fabrication disorder. Fourier spectroscopy provides an effective method to probe even small amounts of scattering and investigate the limits to topological protection and spin-orbit coupling in photonics.
	
	We also quantify the edge-state losses from the reflectance fit. Shaded regions in Fig.~\ref{Fig-crosscut}b depict the fitted linewidths of the modes, which exhibit quality factors ($Q$) of 413 and 486. These deviate not far from the $Q\approx604$ predicted by FEM simulation. The difference could be due to structural disorder, deviation of hole shape with respect to the designed (simulated) structure, or interaction of the propagating mode with radiation fields reflected from the Si substrate. Nonetheless, we conclude that the edge states propagate over distances of many wavelengths in the photonic crystal slab, even though their fundamental Bloch component is inherently phase-matched to the free-space continuum. We expect radiation losses as well as spin-spin scattering to reduce further for smaller deformation of the shrunken and expanded unit cells. However, since that would simultaneously reduce band gap width and increase transverse mode extent, it would be interesting to explore how other strategies, including design of microscopic structure, allow minimizing loss and backscattering.
	
	\subsection*{Imaging propagation and topological routing}
	
	The radiation field also enables mapping edge state propagation in real space, upon locally and selectively exciting them with polarized light. Figure \ref{Fig-realspace}b-d shows images taken with an IR camera using the green beam path in Fig.~\ref{Fig-first}c, when polarized laser light at $1512~\si{\nano\meter}$ is focused at the edge. We observe the topological edge states, propagating out of the excitation focus along the edges between the two lattices. They decay over distances of tens of $\si{\micro\meter}$, in accordance with the measured linewidths. As Figs.~\ref{Fig-realspace}b-d show, right (left) CP light excites only the negative (positive) pseudospin mode which propagates only in the downward (upward) direction. When the edge is excited with vertically polarized light, propagation has equal preference for each direction. 
	
	Finally, we study the routing of topological edge states at sharp corners formed by a junction between an armchair and a zigzag edge that separate four distinct photonic crystal regions, as depicted in Fig.~\ref{Fig-realspace}e. The edges form a total of four topological waveguides emanating from the junction, of two different classes: For negative pseudospin, the two vertical edges support propagation towards the junction whereas the horizontal edges allow propagation away from the junction. For positive pseudospin the directions are reversed.
	
	Figure~\ref{Fig-realspace}f-h show images taken for various excitation points on the vertical edge marked in Fig.~\ref{Fig-realspace}e. We can see in Fig.~\ref{Fig-realspace}f that the negative pseudospin edge state excited with right CP at position 1 propagates downward towards the junction, but not further, as propagation in the bottom waveguide is forbidden for this pseudospin. Had the junction been perfectly symmetric, this pseudospin could have propagated in both the left and right horizontal waveguides. However, the light appears to propagate only into the right waveguide. This is even more clear when the state is excited at position 2, only a few $\mu$m from the junction (Fig.~\ref{Fig-realspace}g). The pronounced observed preference for taking a turn to the right waveguide must be related to the microscopic structural asymmetry of the junction, visible in the inset of Fig.~\ref{Fig-realspace}e. The junction exhibits a chiral character, where we can recognize two continuous interfaces between the expanded and shrunken lattice unit cells that do not cross. Indeed, the light is efficiently routed along the interface that connects the top and right waveguides, even though it is separated from the other interface by sub-unit-cell distance --- much smaller than the transverse extent of the modes. In a control experiment (Fig.~\ref{Fig-realspace}h), we see that the same pseudospin excited in the bottom waveguide follows the other interface, into the left waveguide. Importantly, we also observe neither backreflection nor an enhancement of out-of-plane scattering at the junction. Together, these observations suggest the important role of spin-momentum locking in the routing of the edge states, allowing them to take sharp corners and protect them from scattering to other modes. 
	
	\subsection*{Conclusions}
	To conclude, we have directly excited and observed topologically-protected edge states at telecom wavelengths in a silicon photonic platform. We characterized their propagation, loss, dispersion, and routing along sharp corners. We demonstrated that the pseudospin of the states is inherently encoded in the circular polarization of their far fields, and used this strong spin-orbit coupling to selectively excite the states with a focused laser beam. The employed Fourier spectroscopy technique makes it possible to quantify the inherent spin-spin scattering of the system, opening doors to further understand and control topological protection in QSHE systems, as well as the connections between chirality and topology \cite{fosel_l_2017}. It establishes a straightforward yet versatile path for testing and optimizing novel photonic topological systems for a wide variety of applications including components for integrated photonic chips, quantum optical interfaces, enantiomeric sensing, and lasing at the nanoscale.

	\newpage
	
	\clearpage

	\setcounter{figure}{0}
	\renewcommand{\thefigure}{M\arabic{figure}}
	\setcounter{equation}{0}
	\renewcommand{\theequation}{M\arabic{equation}}

	\section*{Methods}\label{Sec: methods}
	\subsection*{Sample design and FEM simulations}
	
	We used finite element method (FEM) simulations (COMSOL) to optimize the design parameters such that the system can offer topological protection at telecom wavelengths, with both expanded and shrunken lattices exhibiting bandgaps around $1550~\si{\nano\meter}$. A fillet of $25~\si{\nano\meter}$ radius was added to the triangular corners to account for the roundedness of holes in fabrication. The periodicity of the photonic crystal was $800~\si{\nano\meter}$. As indicated in the Supplementary Fig.~S4, equilateral triangles of $250~\si{\nano\meter}$ side were arranged in hexagonal clusters, with the distance from the center of cluster to the center of triangle being $243~\si{\nano\meter}$ for shrunken and $291~\si{\nano\meter}$ for the expanded lattice. Simulations were done for a slab of $220~\si{\nano\meter}$ with a silicon refractive index of 3.48. Perfectly matched layers were added above the simulation box so that simulations of the (quasinormal) eigenmodes also yield their linewidth, which we define as two times the imaginary part of the complex eigenfrequency.
	
	\subsection*{Sample Fabrication}
	The samples were fabricated from a silicon on insulator (SOI) wafer with a $220~\si{\nano\meter}$ Si device layer on top of a $3~\si{\micro\meter}$ buried oxide layer which sits on a $700~\si{\micro\meter}$ thick substrate. A positive tone electron beam (e-beam) resist, AR-P~6209, was spin-coated on top of the waferchips to form a $250~\si{\nano\meter}$ thick resist layer. The photonic crystal designs were patterned in the resist using e-beam lithography (Raith Voyager, 50kV). The resist was developed with Pentyl-acetate/O-xylene/MIBK(9):IPA(1)/IPA. The designed pattern was transferred into the Si device layer using \ce{HBr}/\ce{O2} inductively coupled plasma etching (Oxford Instruments Plasmalab 100 Cobra). Anisole ($5~\si{\minute}$) was used to remove the left-over resist and the sample was cleaned using base piranha (\ce{H2O}(5):\ce{NH4OH}(1):\ce{H2O2}(1)) treatment for 15 minutes at $75\si{\degreeCelsius}$. To obtain free-standing photonic crystal membranes, the oxide layer was wet-etched with HF, followed by critical point drying.
	
	The main sample had an armchair-like edge of $\sim380~\si{\micro\meter}$ length which separates  shrunken and expanded lattices having widths of $\sim196~\si{\micro\meter}$ each. Supplementary Fig.~S1c shows an optical micrograph of this sample. The sample used for the experiment laid out in \cref{Fig-realspace}e  contained two armchair-like edges of $\sim55~\si{\micro\meter}$ and two zigzag-like edges of $\sim80~\si{\micro\meter}$ forming a four-way junction separating two sets of expanded and shrunken lattices.

	\subsection*{Experimental Details}
	\subsubsection{Dispersion Measurements}
	A 200mW supercontinuum source (Fianium WhiteLase Micro) generates light with a broadband spectrum. Its output is filtered by a long pass filter with cutoff wavelength 1150 nm and coupled into a single mode optical fiber. The IR light from the fiber is collimated by an achromatic lens and passed through an OD:1 neutral density filter, linear polarizer (LP) and an achromatic quarter-wave plate (QWP), which together define the polarization of the input beam. A beamsplitter (BS) steers the input light to an aspheric objective (Newport 5721-C-H, 60x, NA=0.6) which focuses light on the sample. Reflected light is collected by the same objective and passed through the beamsplitter to the Fourier lens (FL) placed in the output path. This lens, together with a tubelens (TL), images the objective back focal plane (BFP) onto the entrance slit of a spectrometer (Acton SpectraPro SP-2300i). Optional custom spatial filters (SF) are placed in the image plane between the FL and TL to define the sample area from which light is collected. The (vertical) entrance slit of the spectrometer, aligned with the optical axis, chooses $k_x=0$ in the reciprocal plane, confirmed using a test grating sample. With the help of two parabolic mirrors (PM) for focusing and collection, the spectrometer grating then disperses the broadband IR light orthogonal to the slit, such that the InGaAs IR camera (AVT Goldeye G-008 SWIR) placed at the spectrometer output records images of frequency versus $k_y/k_0$, where $k_0$ is the free-space wavevector. The spectral resolution is $81.5~\si{GHz}$, and the wavevector resolution $\Delta k_y/k_0\approx0.011$. The wavevector axis was calibrated by observing the diffraction of a gold transmission grating with known pitch placed at the sample position. The frequency axis was calibrated using laser sources of known wavelength. Reflectometry images taken at different center frequencies of the grating are stitched together in post-processing to plot dispersion over a wide range of frequencies.
	\subsubsection{Polarimetry Measurements}	
	We use the ``rotating QWP'' scheme outlined in \cite{schaefer_measuring_2007} to measure the Stokes polarization parameters. We place a QWP and LP after the FL so that the reflected beam passes through them sequentially. For a given rotation angle $\ta$ of the QWP, the detected intensity can be written in the form of a truncated Fourier series
	\begin{equation}\label{Eq: IAB}
	I(\ta)=\frac{1}{2}(A+B\sin 2\ta+C\cos 4\ta+D\sin 4\ta),
	\end{equation}
	where the Fourier coefficients A, B, C and D are related to the measured Stokes parameters as $S'_{0}=A-C$, $S'_{1}=2C$, $S'_{2}=2D$ and $S'_{3}=B$. For our polarimetry measurements, we take 12 successive measurements rotating the QWP from $\ta_{1}=0\si{\degr}$ to $\ta_{12}=165\si{\degr}$ in steps of $15\si{\degr}$ to find these coefficients as:
	\begin{equation}\label{Eq: IQWP}
	\begin{aligned}
	A &= \frac{2}{12}\sum_{n=1}^{12}I(\ta_{n}),& 
	B &= \frac{4}{12}\sum_{n=1}^{12}I(\ta_{n})\sin 2\ta_{n},\\
	C &= \frac{4}{12}\sum_{n=1}^{12}I(\ta_{n})\cos 4\ta_{n},&
	D &= \frac{4}{12}\sum_{n=1}^{12}I(\ta_{n})\sin 4\ta_{n}.
	\end{aligned}
	\end{equation}
	The LP after the QWP is fixed at $\si{0\degr}$ with respect to the $\hat{x}$ axis in all these measurements. 
	
	The collected light is not fully polarized. The polarized component for a given output is defined as $I_p = \sqrt{S_1'^2+S_2'^2+S_3'^2}$. We normalize the measured Stokes parameters to the maximum value of the polarized component in the output as $S_i = S'_i/\max(I_p)$ for $i=1,2,3$.
	Results of full polarization tomography of the edge states is given in Supplementary Fig.~S5.

	Contributions from background reflections can be important while studying the Stokes parameters of a resonant mode’s radiation, as the observed polarization of the superposition of both fields can depend strongly on their (uncontrollable) relative phase. We isolate the edge-mode radiation in the far field from the direct reflection components by altering the SF. Supplementary Fig.~S6 shows a sketch of the employed SFs, determining the collection area during edge state dispersion measurements without and with a center-block. With the latter, collection of direct reflections from the illumination spot is prevented, but radiation from edge states that propagate out of the center spot is detected.
	
	\subsubsection{Real-space imaging}
	To visualize the propagation of edge modes in real-space, we take out the FL, SF and polarimetry optics in the output path. Fiber-coupled light at $1512~\si{\nano\meter}$ from a single wavelength CW tunable laser (Toptica CTL) is delivered through the same IR fiber. The QWP and LP in the input path control polarization. In this case, the collected light is steered to a different path after the BS and the sample is imaged onto an IR camera (AVT Goldeye P-008 SWIR) using another tube lens.
	
	\subsection*{Extraction of modes from dispersion crosscut}
	To the normalized reflection obtained from the cross-cut in Fig.~\ref{Fig-crosscut}a, we fit a set of complex Lorentzians of the form:
	\begin{equation}
	R(\oo)=\left|A_{0}+\sum_{j=1}^{3}A_{j}\ee^{\ii \phi_{j}}\frac{\gamma_j}{\oo-\oo_{{0}_{j}}+\ii\gamma_j}\right|^2,
	\end{equation}
	where $A_{0}$ is a constant background amplitude; $A$, $\phi$ and $\oo_{{0}}\!-\ii\gamma$ are the amplitude, phase and complex frequency of three individual Lorentzians. Quality factors are defined as $Q_j=\oo_{{0}_{j}}/(2\gamma_j)$. Two of these Lorentzians model the edge modes, while the third (broad) Lorentzian accounts for the slowly varying background reflection.
	
		\section*{acknowledgements}
	N.P. thanks H.M. Doeleman and T.A. Bauer for their suggestions and help in setting up the Fourier microscope. The authors thank S.R.K. Rodriguez for critical reading of the manuscript. This work is part of the research programme of the Netherlands Organisation for Scientific Research (NWO). The authors acknowledge support from an industrial partnership between Philips and NWO, and from the European Research Council (ERC Advanced Grant No. 340438-CONSTANS and ERC Starting Grant No. 759644-TOPP) and the Marie Sk\l odowska-Curie Actions (individual fellowship BISTRO-LIGHT, No. 748950). 
	
	\section*{Author contributions}
	N.P. carried out the simulations, device fabrication, experiments and analysis. F.A. developed the theoretical model and assisted with simulations. E.V. and L.K. supervised the work. All authors contributed extensively to the project conception, interpretation of results, and writing of the manuscript.
	
	\newpage
	\clearpage

	\section*{Supplementary Information}

	\setcounter{figure}{0}
	\renewcommand{\thefigure}{S\arabic{figure}}
	\setcounter{equation}{0}
	\renewcommand{\theequation}{S\arabic{equation}}

		\subsection*{Direct Fabry-P\'erot reflections and reflectometry of an ordinary lattice}
	Here we comment on the broad low-reflection fringes in our dispersion measurements. The $3~\si{\micro\meter}$ air cavity formed between the suspended photonic crystal and the Si substrate can effectively act as a Fabry-P\'erot cavity and can contribute to the measured reflected intensity of the sample. 
	\begin{figure}[!h]
		\centering
		\includegraphics{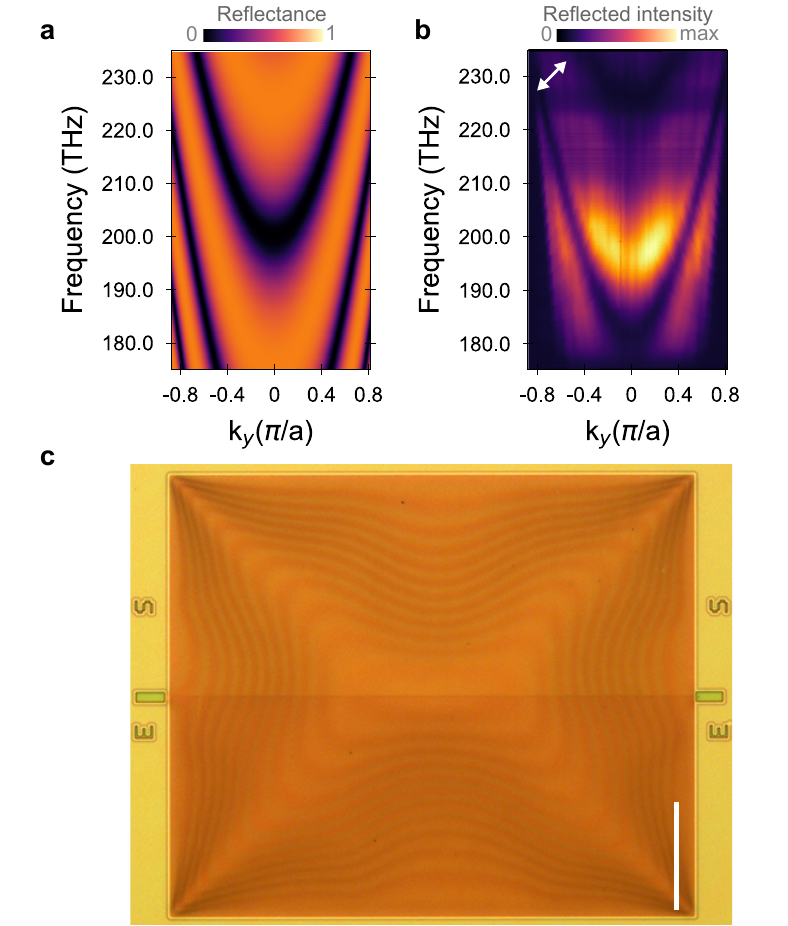}
		\caption{a. Analytically calculated Fabry-P\'erot-like fringes for an air cavity between two Si layers. b. Measured dispersion for an ordinary honey-comb-like lattice for diagonally polarized incidence. c. Optical micrograph of the fabricated sample. Scalebar corresponds to 100 $\si{\micro\meter}$ on the sample.}
		\label{Fig-suppl-FP}
	\end{figure}
	To confirm this, we analytically calculate the reflection of an aircavity between two Si surfaces. As shown in \cref{Fig-suppl-FP}a, the resulting parabolic dispersion curves are comparable to what we observe in the reflection measurements shown in the main text. The position of these curves shift in frequency as a function of height of the air layer. In our sample, the height of air cavity below the suspended membrane is different at different points in the photonic crystal. We see this buckling from the color contrast in the optical micrograph as shown in \cref{Fig-suppl-FP}c. This results in spectral shifting of the low-reflection fringes when we take dispersion measurements at expanded, shrunken and edge locations.

	In addition to the expanded and shrunken lattices, we also fabricate a photonic crystal without any symmetry breaking, i.e., triangular holes arranged in a perfect honeycomb lattice. 
	
	In such a system, there are no photonic crystal modes that can couple to the far-field: Whereas the dispersion features in principle a doubly degenerate Dirac cone when considering a six-site hexagonal unit cell, the Bloch states carry no weight in the fundamental harmonic due to the lattice symmetry. In \cref{Fig-suppl-FP}b we show the measured reflected intensity for a honeycomb lattice of triangular holes. As expected, we don't see any sharp resonances in the spectra except for the background fringes from the Fabry-P\'erot-like air cavity.
	
	\subsection*{Orthogonal linear polarization of bulk modes}
	
	\begin{figure}[!h]
		\centering
		\includegraphics[width=1\columnwidth]{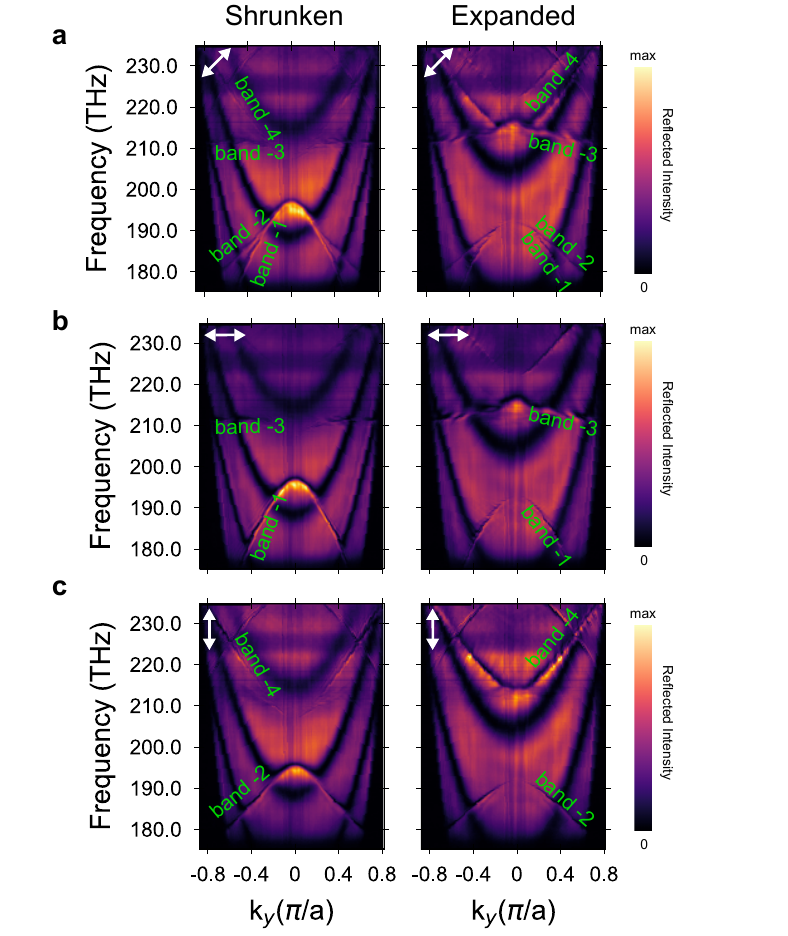}
		\caption{Dispersion of bulk modes for orthogonally polarized incidences. a-c: Far-field reflection as a function of $k_y/k_0$ at shrunken and expanded lattices excited with light that is linearly polarized along $\si{\ang{45}}$ with respect to $\hat{x}$ (a), along $\hat{x}$ (b), and along $\hat{y}$  (c). The four linearly polarized bulk modes are marked as band-1 to band-4 in a.}
		\label{Fig-bulklp}
	\end{figure}
	Here we observe that the bulk modes are linearly polarized with orthogonal polarizations in the far-field. They can be selectively excited with vertical and horizontal polarizations. 
	
	Figure~\ref{Fig-bulklp} shows photonic band dispersion in shrunken and expanded lattices when those are excited with light of different linear polarizations. The top panel of Fig.~1d of the main text is reproduced here as Fig.~\ref{Fig-bulklp}a,  and shows the reflected light intensities as a function of the reciprocal lattice vector $k_y$ when diagonally polarized light is incident on the sample. Here the bandgap region extends from $197~\si{\tera\hertz}$ to $208~\si{\tera\hertz}$. Let us have a closer look at the two bands (marked as band-1 and band-2 in the figure) directly below the gap for shrunken and expanded lattices. These are bright modes for the shrunken and dark modes for the expanded lattices. Now, if we excite the sample with horizontally polarized light, we can see that the incident light excites only one of these two modes (band-1) as shown in Fig.~\ref{Fig-bulklp}b. However, for vertically polarized incidence, it is the other mode (band-2) that is getting excited as we see in Fig.~\ref{Fig-bulklp}c. This suggests that band-1 is horizontally polarized and band-2 is vertically polarized. Thus, the two bands below the gap are linearly polarized orthogonal to each other. 
	
	We see a similar behaviour also in the bands that are located above the gap. As we see in Fig.~\ref{Fig-bulklp}a, band-3 is excited for horizontally polarized incidence. And for vertically polarized incidence, band-4 is excited as can be seen from Fig.~\ref{Fig-bulklp}c.
	\subsection*{Tight-binding model}
	Here we extend the tight binding model outlined in Ref.~\cite{barik_two-dimensionally_2016} to support our experimental observations. In the existing model, the Hamiltonian of the system is calculated with a nearest neighbour approximation making a linear approximation in reciprocal lattice vectors $k_x$ and $k_y$. The model also restricts the six-dimensional basis to a four-dimensional system. These two assumptions, however, result in the prediction of fully degenerate bulk states and gapless edge states. We expand the theory by going beyond these assumptions.
	
	We consider a six-dimensional vector $\bk{\ps}$ where each component represents one of the six sites in a hexagonal unit cell. We observe that in order to get results in agreement with the experimental results, it is required to include next-neighbour interaction up to third order in the model. The tight-binding Hamiltonian reads:
	\begin{widetext}
		\begin{equation}\label{Eq: 6x6h}
		H = -\left(\begin{array}{cccccc}
		0 & \intra1 + \inter3\expt{(\av1-\av2)} & \intra2 + W_1  & \intra3 + \inter1\expt{\av1}  & \intra2 + W_5 & \intra1 + \inter3\expt{\av2} \\
		0 & 0 & \intra1 + \inter3\expt{\av1} & \intra2 + W_2 & \intra3 + \inter1\expt{\av2} & \intra2 + W_6 \\
		0 & 0 & 0 & \intra1 + \inter3\expt{\av2} & \intra2 + W_3 & \intra3 + \inter1\expt{(\av2-\av1)} \\
		0 & 0 & 0 & 0 & \intra1 + \inter3\expt{(\av2-\av1)} & \intra2 + W_4 \\
		0 & 0 & 0 & 0 & 0 & \intra1 + \inter3\expt[-]{\av1} \\
		0 & 0 & 0 & 0 & 0 & 0 \\
		\end{array}\right) + \text{H. c.},
		\end{equation}
	\end{widetext}
	where, for the sake of compactness, we define the auxiliary vector
	\begin{multline}
	\vec{W} = \inter2 (\expt{\av1} + \expt{(\av1-\av2)}, \expt{\av1} + \expt{\av2},\expt{\av2} + \expt{(\av2-\av1)},\\
	\expt[-]{\av1} + \expt{(\av2-\av1)},\expt{\av1} + \expt{\av2},\expt{\av2} + \expt{(\av2-\av1)})^T,
	\end{multline}
	
	and the lattice unit vectors $\av1 = (a,0)^T$ and $\av2 = a/2\,(1,\sqrt{3})^T$. We also designate the intra-cell coupling coefficients for first-, second-, and third-nearest neighbors as $\intra1,\intra2,$ and $\intra3$, respectively. Similarly, we indicate with $\inter1, \inter2$, and $\inter3$ the inter-cell coupling coefficients.
	
	The modes in the photonic crystal are mostly TE polarized. Therefore, we assume that the six components of $\bk{\ps}$ correspond to the field component $H_z$ at the six lattice sites.
	For a TE mode, the in-plane electric field components are proportional to:
	\begin{equation}
	E_x \propto -\frac{\partial H_z}{\partial y}, \quad E_y \propto \frac{\partial H_z}{\partial x}.
	\end{equation}
	We assume that inside each hexagonal unit cell, the field varies mostly along the angular dimension $\theta$, i.e., $H_z \simeq H_z(\theta)$ and that the variation of $H_z$ between one site and another is such that the variable can be considered continuous. Thus, moving to cylindrical coordinates, the electric field components become:
	\begin{equation}
	E_x \simeq \cos(\theta) \frac{\partial H_z}{\partial \theta}, \quad E_y \simeq \sin(\theta) \frac{\partial H_z}{\partial \theta},
	\end{equation}
	so that we can write $\vec{E} \simeq \hat{\vec{R}}\:\partial_{\theta} H_z$, where $\hat{\vec{R}}$ is the radial unit vector from the center of the unit cell.
	
	The polarization density is proportional to the electric field: $\vec{P} = \chi \vec{E}$. We are interested in the total dipole moment inside the unit cell
	\begin{equation}
	\vec{p} = \int_{\mathrm{cell}} d\vec{r}\,  \vec{P}(\vec{r}).
	\end{equation}
	Using the previous assumptions, we can approximate the total dipole moment as follows:
	\begin{equation}
	\vec{p} \propto \int_{0}^{2\pi} d\theta \hat{\vec{R}}(\theta) \frac{\partial H_z}{\partial \theta}.
	\end{equation}
	By partial integration, this becomes
	\begin{equation}
	\vec{p} \propto \int_{0}^{2\pi} d\theta \frac{\partial \hat{R}(\theta)}{\partial \theta} H_z(\theta) \propto \int_{0}^{2\pi} d\theta  \,\hat{\vec{\theta}}\, H_z(\theta).
	\end{equation}
	where $\hat{\vec{\theta}}$ is the unit vector along the azimuthal direction. 
	
	Now we can go back to the discretized system and replace the integral over $\theta$ with the sum over the six sites of the hexagon. In this way, we obtain the following estimation for the total dipole moment inside the unit cell
	\begin{equation}\label{Eq: celldipole}
	\vec{p} \propto \sum_{i=1}^6 \hat{\vec{\theta}}_i \psi_i,
	\end{equation}
	where $\hat{\theta}_i$ is the azimuthal unit vector of the $i$ site in the unit cell, assuming the origin at the center of the hexagon. In practice, $\hat{\theta}_i = (-\sin[(i-1)\pi/3],\cos[(i-1)\pi/3])^T$, with $i = 1,\dots,6$.
	
	\begin{figure}[!h]
		\centering
		\includegraphics[width=1\columnwidth]{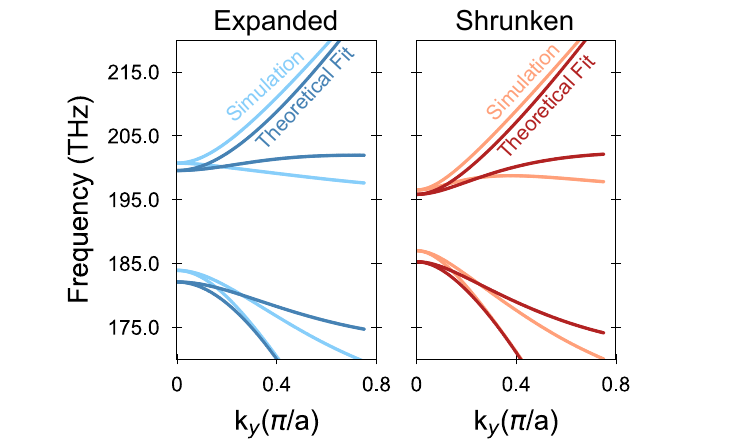}
		\caption{Comparison of bandstructure calculations for bulk lattices: FEM vs tight-binding model. The lighter lines correspond to dispersion calculated by FEM simulation whereas the darker lines correspond to dispersion predicted analytically by our tight-binding model. Left panel shows dispersion in expanded lattice and the right panel shows dispersion in shrunken lattice.}
		\label{Fig-theory-fit}
	\end{figure}
	We fitted the parameters $t_1, t_2, t_3$ and $s_1, s_2, s_3$ of the Hamiltonian to reproduce the bandstructure calculated with the finite-element method for both the expanded and the reduced crystals. In Fig. \ref{Fig-theory-fit}, we compare the FEM results with the tight-binding method. We observe that including higher-order-interaction terms into the tight-binding Hamiltonian is essential for correctly reproducing the splitting of the bands away from the $\Gamma$ point. Moreover, the dipole moment calculated from Eq.~\eqref{Eq: celldipole} confirms that the bulk modes are linearly polarized in the far field, in full agreement with our experiment.

	We can also use the tight-binding model to calculate the edge states. In order to do so, we consider the following system: two regions of space, $A$ and $B$, separated by an edge at $x = 0$. In regions $A$ and $B$ we have the shrunken and expanded crystals with the Hamiltonians $H_A$ and $H_B$, respectively. Both Hamiltonians have the form in Eq.~\eqref{Eq: 6x6h}, but each with a different set of parameters. Each Hamiltonian is a function of $k_x$ and $k_y$. Since the model is still translation-invariant along $y$, $k_y$ is conserved. However, $k_x$ is not conserved and, therefore, needs to be replaced with a corresponding differential operator, according to the substitution $k_x = -i\partial_x$. For simplicity, we expand both Hamiltonians $H_{A}(k_x,k_y)$ and $H_{B}(k_x,k_y)$ to first order in Taylor series
	\begin{align}
	H_{A,B} \simeq &~H_{A,B}(\vec{0})\nonumber\\ &~+
	k_x \left.\frac{\partial H_{A,B}}{\partial k_x}\right|_{k_x=0, k_y = 0}\nonumber\\ &~+ k_y \left.\frac{\partial H_{A,B}}{\partial k_y}\right|_{k_x=0, k_y = 0}.
	\end{align}
	
	For the edge states, we consider only the frequencies inside the bandgap of the bulk modes. Therefore, there are no propagating solutions for the Schr\"{o}dinger equation of the Hamiltonians $H_A$ and $H_B$. However, there are evanescent solutions, which are defined by the (generalized) eigenvalue problems for $k_x = i\delta^{(A)}$ and $k_x = i\delta^{(B)}$ in the two regions separately:
	\begin{align}
	\left[\omega - H_{A}(\vec{0}) - k_y \left.\frac{\partial H_{A}}{\partial k_y}\right|_{0}\right]
	\bk{\psi_m^{(A)}} =& i \delta_m^{(A)}
	\left.\frac{\partial H_{A}}{\partial k_x}\right|_{0}\bk{\psi_m^{(A)}},\\
	\left[\omega - H_{B}(\vec{0}) - k_y \left.\frac{\partial H_{B}}{\partial k_y}\right|_{0}\right] \bk{\psi_m^{(B)}} =& i \delta_m^{(B)}
	\left.\frac{\partial H_{B}}{\partial k_x}\right|_{0}\bk{\psi_m^{(B)}}.
	\end{align}
	Here, the index $m$ runs over the different eigenvalues and eigenvectors of the two equations. These eigenvalue equations imply that the solutions of the Schr\"odinger equation for $\bk{\ps}$ can be written as 
	\begin{equation}
	\bk{\ps(x)} = \sum_{m = 1}^{3} \alpha_n^{(A)} \bk{\psi_{m}^{(A)}} \exp(|\delta_m^{(A)}|x) 
	\end{equation}
	in region $A$ (say, $x < 0$), and
	\begin{equation}
	\bk{\psi(x)} = \sum_{m = 1}^{3} \alpha_m^{(B)} \bk{\psi_{m}^{(B)}} \exp(-|\delta_m^{(B)}|x)
	\end{equation} 
	in region $B$ ($x > 0$). In both cases, $\alpha_{m}^{(A,B)}$ are expansion coefficients.
	
	In order to find the frequencies of the edge states, we have to enforce a continuity condition at the point $x = 0$:
	\begin{equation}
	\sum_{m = 1}^{3} \alpha_m^{(A)} \bk{\psi_{m}^{(A)}} = 
	\sum_{m = 1}^{3} \alpha_m^{(B)} \bk{\psi_{m}^{(B)}}.
	\end{equation}
	This condition leads to the characteristic equation for the determinant of a $6\times 6$ matrix, which we solve numerically. The solutions of the characteristic equation correspond to the frequencies of the edge states for a given value of $k_y$.
	
	Moreover, from the coefficients $\alpha_{m}^{(A)}$ and $\alpha_{m}^{(B)}$ we can reconstruct the profile of the edge state along the $x$ axis. Using Eq.~\eqref{Eq: celldipole}, we can compute the local cell-averaged dipole moment, $\vec{p}(x)$. Finally, the total dipole moment of the edge state is calculated by integrating along $x$, i.e.,
	\begin{equation}
	\vec{p}_{\text{tot}} = \int dx\: \vec{p}(x).
	\end{equation}
	The frequencies of the edge states calculated with this method and the $S_3$ parameter derived from the corresponding total dipole moments are shown in Fig.~2d in the main text.

	\subsection*{Sample design parameters}
	\begin{figure}[!h]
		\centering
		\includegraphics[width=.7\columnwidth]{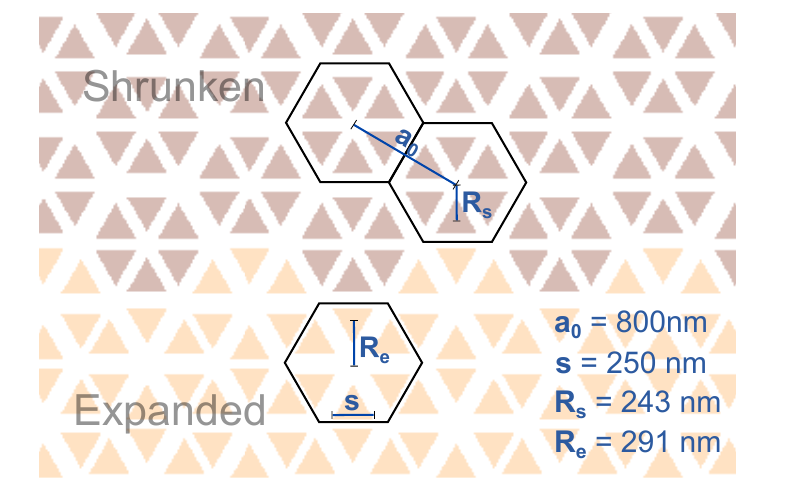}
		\caption{Design parameters for shrunken and expanded lattices. The unit cell period $a_0=800~\si{\nano\meter}$ and the length of the triangle sides $s$ are the same for both lattices. The distance from the center of the unit cell to the center of the triangle was different for shrunken ($R_s=243~\si{\nano\meter}$) and expanded ($R_e=291~\si{\nano\meter}$) lattices.}
		\label{Fig-geoemetry}
	\end{figure}
	
	Figure~\ref{Fig-geoemetry} shows different geometric parameters of the simulated system as explained in Methods. 
	
	\subsection*{Polarization tomography of edge states}
	
	\begin{figure}[!h]
		\centering
		\includegraphics{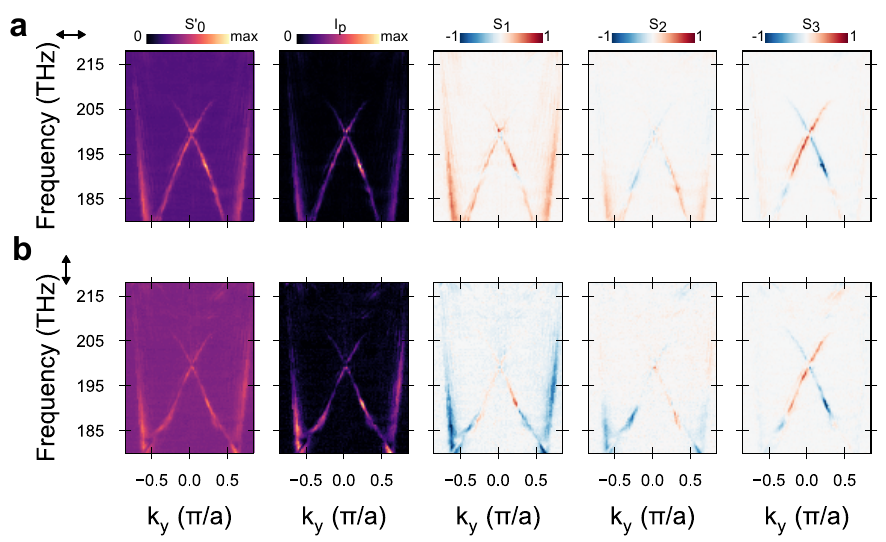}
		\caption{Polarization tomography of the edge states. Intensity of the polarized reflected field and various Stokes parameters of the edge states for incident light linearly polarized along $\hat{x}$ (a) and along $\hat{y}$ (b).}
		\label{Fig-poltomo}
	\end{figure}
	
	For the sake of completeness, we show the full polarization tomography of the edge states for horizontally and vertically polarized incidences in \cref{Fig-poltomo}.

	\Cref{Fig-poltomo} shows the measured Stokes parameter $S'_0$, intensity of the polarized component in the reflected signal $I_p = \sqrt{S_1'^2+S_2'^2+S_3'^2}$, and normalized values $S_1$, $S_2$ and $S_3$ (normalized to the maximum value of polarized component $I_p$ as mentioned in  Methods).

	\subsection*{Spatial filtering}

\begin{figure}[!h]
		\centering
		\includegraphics{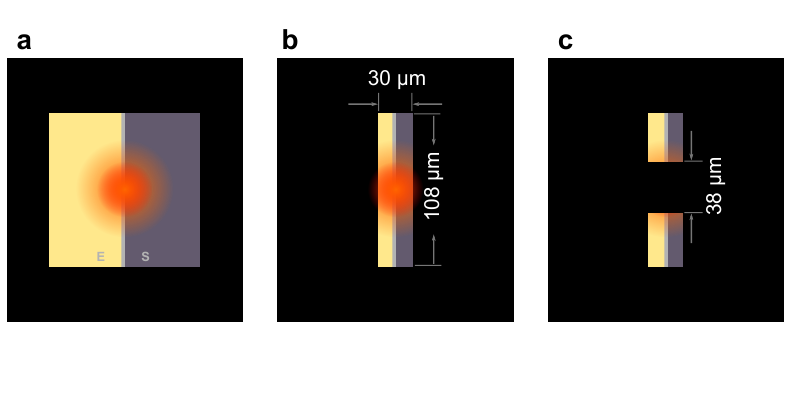}
		\caption{A cartoon depicting modification of the spatial filter placed in a real-space image plane in the detection path to isolate contribution from the edge modes for S3 measurement. a. Real-space plane without the slit. The illumination spot is indicated in orange. b. The slit defines the sample area from which light is collected. b. Putting a centerblock across the slit blocks most of the illumination spot, allowing light to pass originating only from the edges. Dimensions of the effective collection area at the sample as well as the dimension of the blocked region are marked along the sides.}
		\label{Fig-centerblock}
\end{figure}

	\Cref{Fig-centerblock}a and b show schematic representations of the collection area at the sample in real space without and with the spatial filter respectively.
	
		 For polarimetry measurements on the edge, the collection area was modified by putting a block at the center of the slit as shown in \cref{Fig-centerblock}c.


\newpage\clearpage
\end{document}